# BlackHolistic 2026 meeting report

*Accretion and relativistic jet formation take place across the black hole mass range, from black holes of just a few solar masses to those in excess of ten billion. Despite the enormous range in scales, qualitative similarities and quantitative scalings appear to connect the entire population. In March 2026, researchers from across the black hole mass spectrum met in Oxford to educate, explore and forge new research directions.*

**Rob Fender[1], Jane Dai[2], Erin Kara[3], Sera Markoff[4,5], Francesca Panessa[6], Jiri Svoboda[7], and Heino Falcke[8].**

1 University of Oxford, Oxford, UK.
2 University of Hong Kong, Pok Fu Lam, Hong Kong SAR, China.
3 Massachusetts Institute of Technology, Cambridge, MA, USA.
4 University of Amsterdam, Amsterdam, Netherlands.
5 University of Cambridge, Cambridge, UK.
6 Istituto di Astrofisica e Planetologia Spaziali (INAF), Rome, Italy.
7 Astronomical Institute of the Academy of Sciences, Prague, Czech Republic.
8 Radboud University, Nijmegen, Netherlands.

Corresponding author e-mail: rob.fender@physics.ox.ac.uk

**A zoo of black hole accretors and the road to unification**

Black holes are the most exotic objects in our Universe. When they accrete matter close to their event horizons, they release huge amounts of radiative and kinetic energy. The kinetic energy output often takes the form of highly collimated outflows moving at relativistic speeds, which can sometimes extend to enormous scales (many times the sizes of entire galaxies), which we call ‘jets’. Relatively nearby and low mass, black holes of a few solar masses in X-ray binary systems in our Galaxy have revealed patterns of behaviour, which tell us about the temperature, density and flickering of the accretion flow. These patterns are called ‘accretion states’ and they demonstrate a clear and repeatable connection between accretion and jet formation. Comparison with populations of the most supermassive black holes, weighing in at a billion solar masses or more and largely frozen in time from a human viewpoint, suggests that similar patterns may exist, a range of over 100 million in scale. Between these two extremes, the recently established classes of Tidal Disruption Events (TDEs) and Changing-Look Active Galactic Nuclei (AGNs), which are highly variable despite having black holes in excess of a million solar masses, suggest that in some cases we may be able to test, in near real time, whether accretion and jet formation in black holes proceeds in similar ways across the entire mass spectrum.

In the early 2000s, when X-ray accretion states and their connection to jets were being established for stellar mass black holes, there was great interest in whether accretion unification across the black hole mass range could be achieved. Early successes included the scaling of radio and X-ray luminosities with black hole mass dependence, to produce a 'fundamental plane of black hole activity', and apparently X-ray-binary-like patterns of state change and jet ejections in supermassive black holes like 3C120.

In recent years this unification has been the goal of the SHIVA (*Search for Holistic Interpretation of Variable Accretion*) series of workshops. Inspired by SHIVA, and motivated (and funded) by the 'Blackholistic' ERC Synergy grant, which had just been awarded to three of us (Falcke, Fender & Markoff), with the express goal to test unification of black hole jet production, we organised this meeting to bring these communities back together. The meeting was an enormous success, three times oversubscribed for places (which translates to nearly 400 abstracts submitted), and proved to be extremely stimulating for all participants.

**Towards and beyond scale-free accretion states**

It was known by the 1990s that black hole X-ray binaries, which underwent spectacular outburst typically lasting a few months, cycled through a number of distinct X-ray states. These were initially characterised by X-ray spectral variability and later, arguably more precisely, by temporal X-ray fluctuations.

A major question at the conference was whether such states exist in black holes of all masses. Three highlight talks discussed the state of the art in our understanding for X-ray binaries (Motta), TDEs (Alexander) and Changing-Look AGNs (Ricci). Many contributed presentations tried to directly address the question of whether these accretion states are really scale-free or not, and we encourage the readers to visit our talks and posters archive via the links at the end of this article. Cutting-edge observations demonstrate how advances in the field remain driven by new facilities (such as GRAVITY, EHT), well-coordinated observational campaigns, and serendipity.

A clear consensus was hard to achieve: the similarities seem inescapable (steady/steadier jets at lower Eddington fraction accretion rates, discrete ejecta only at the highest accretion rates, the association of these ejections with 'state' changes), but the next steps in unification remain elusive.

Beyond the accretion states observed in unobscured accreting black holes, fresh momentum was given to the discussion of how dense matter very close to the black hole can affect how we observe accretion and the behaviour of the relativistic jet. Are obscured, super-Eddington accretors in the local universe, such as SS433, related to the 'Little Red Dots' now being discovered at high redshift? This was reviewed by Webb and discussed in many contributed presentations. Jets propagating into dense environments such as those of the obscured/super-Eddington sources may also have a connection to the detections of very high gamma ray emission detected in recent years from X-ray binaries, as reviewed in the highlight talk of Olivera-Nieto.

**Black hole anatomy in the eyes of simulations**

Numerical simulations of accretion, jet formation and indeed the entire evolution of the universe were reviewed (Davis, Hopkins). While all scales can now be simulated, no single simulation is able to capture them all, and the interface between simulations and observations remains an exciting fruitful landscape. A good example of this is horizon-scale simulations of black hole accretion and jet formation, which seem to require a rapidly rotating black hole to produce the fastest and most powerful jets, whereas there remain relatively slowly spinning objects (e.g. some accreting neutron stars) which also seem to be able to do similar things. The horizon-scale simulations are now being used to directly infer physical properties of the accretion flows which are directly imaged by the Event Horizon Telescope, such as their density and magnetisation.

Once it was established, in the past decade, that numerical simulations of structure formation and galaxy growth required strong kinetic and radiative feedback from central supermassive black holes to regulate the growth of the most massive galaxies, the question of how to implement this feedback into the models has been a hot topic. This was the subject of the highlight talk by Sijacki on the final morning, and it connects larger scales, for those of us who work on accretion states and jets in detail, to smaller scales.

In the final talk of the meeting, Beckmann discussed the schemes for implementing states as a function of accretion rate, with associated prescriptions for radiation and kinetic (jet) feedback, into these simulations. She also showed how one may make assumptions about the relationship between AGN jet power and black hole spin, combine these with assumptions about whether the BH is spinning up or down in different states, and use simulations and observations to infer the spin evolution of the most massive black holes over cosmological time.

**A strong format and an archive of talks**

Beyond the array of excellent talks, a strength of the conference was the  well prepared interactive discussion sessions, which took place in the early afternoon each day. Carefully chosen panels curated these hour-long sessions, including taking questions both online via the conference slack and in person in the room. Every discussion remained lively and highly interactive for the full hour, and generated many discussions which continued.

The discussion sessions were:

- Monday: *How well do simulations capture reality?* (Davis, Markoff, Matthews, Rossi)

- Tuesday: *Does spin power black hole jets?* (Beckmann, Done, Fabian, Fender)

- Wednesday: *Are the same accretion states and outburst cycles present in all black holes?* (Motta, Ricci, Svoboda, Uttley)

- Thursday: *Super-Eddington flows – identifying universal signatures* (Begelman, Dai, Middleton)

- Friday: *Are we putting the right black hole physics in cosmological simulations?* (McNamara, Merloni, Panessa, Sijacki)

In addition to these discussions, we had a small number of non-traditional sessions at the conference in order to break up the astrophysics and provide some different stimulation. These included a 'speed networking' session, a discussion on the use of AI in Astrophysics (featuring contributions from Bowles and Stewart), a stimulating review on neurodiversity and its role in science (Tsompanidis) and a chance to meet and interact with four calming animals. We had two public outreach events (Tuesday and Thursday evenings) during the week and were delighted to have Sir Roger Penrose in attendance for much of the meeting.

**Takeaway**

The meeting highlights the power of connecting studies of black holes in X-ray binaries and AGN, where comparing similarities and differences across scales is key to advancing black-hole accretion physics. Strengthening the link between observations and simulations remains essential, as does continued interaction between communities that have too often evolved separately. The strong international presence and active engagement of students and other early-career researchers highlighted a vibrant and growing field, reinforcing the importance of future meetings that bring these diverse perspectives together.

All of the talks, discussions, and most of the posters at the meeting were recorded and are publicly available. You can find these resources at the following link, and we encourage you to visit them:

https://blackholesoxford2026.web.ox.ac.uk/programme

---

**Acknowledgements**

The meeting was supported largely via European Research Council Synergy Grant 101071643 'Blackholistic', which has principal investigators Heino Falcke, Rob Fender and Sera Markoff.

We are grateful to local University of Oxford volunteers who helped enormous with the logistics of the meeting (Leanne O'Donnell, Ashling Gordon, Lawrence Selmons-Clark, Alex Cooper, Clara Lilje, Emma Elley, Fraser Cowie, Hengyue Zhang, James Matthews, Katie Savard, Steve Prabu, Wei-Bo Kao, Zuobin Zhang) as well as additional help from Maya Garbaccio Gili (Rome). The alpacas were enthusiastically organised by Clara Lilje.

**Competing interests** The authors declare no competing interests.

**Figure 1.** Blackholistic 2026 conference attendees standing on the Penrose paving outside the Mathematical Institute, University of Oxford, which hosted the meeting. We thank Emma Elley for revealing the black hole.

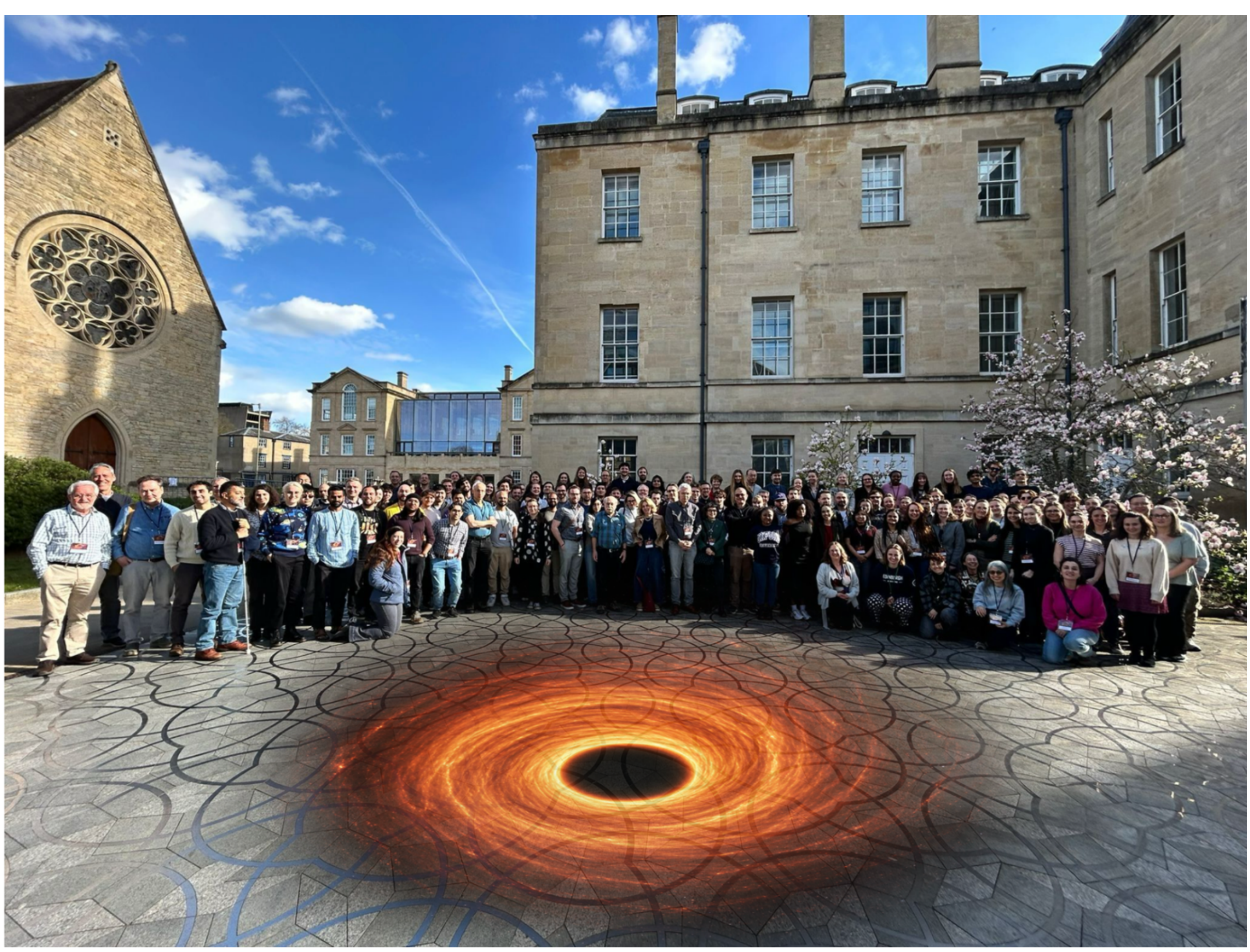